\documentclass[useAMS,usenatbib]{mn2e}
\usepackage{times}
\usepackage{graphicx}
\usepackage{aas_macros}

\newcommand{\msun}{\mbox{${\rm M}_{\odot}$ }}
\def\rev#1{ #1}

\title[AM~CVn systems as optical, X-ray and GWR sources]{
Short-period AM\,CVn systems as
  optical, X-ray and gravitational wave sources}

\author[G. Nelemans et al.]  {G. Nelemans$^{1}$\thanks{E-mail:
    nelemans@ast.cam.ac.uk} and L. R. Yungelson$^{2,3}$ and S.F.  Portegies Zwart$^{4}$ 
\\
  $^{1}$Institute of Astronomy, University of Cambridge, Madingley Road, Cambridge CB3 0HA, UK\\
  $^{2}$Institute of Astronomy of the Russian Academy of
  Sciences, 48 Pyatnitskaya Str., 109017 Moscow, Russia \\
  $^{3}$Isaac Newton Institute, Moscow Branch, 13 Universitetskii pr.,
  119899, Moscow, Russia \\
  $^{4}$Astronomical Institute ``Anton Pannekoek'' and Section
  Computational Science, University of Amsterdam,
  Kruislaan 403, NL-1098 SJ Amsterdam, the Netherlands\\
}

\defcitealias{npv+00}{NPVY}

\begin{document}

\date{Accepted . Received \today}

\pagerange{\pageref{firstpage}--\pageref{lastpage}} \pubyear{2002}

\maketitle

\label{firstpage}

\begin{abstract}
  We model the population of AM~CVn systems in the Galaxy and discuss
  the detectability of these systems with optical, X-ray and
  gravitational wave detectors.  We concentrate on the short period
  ($P <$ 1500 s) systems, some of which are expected to be
  in a phase of direct impact accretion.
%We develop simple models for the
%  emission of optical and X-ray radiation from the AM~CVn stars. 
%Using
%  a self-consistent model for the star formation history and radial
%  distribution of stars in the Galaxy plus a simple model for
%  interstellar absorption in optical and X-rays, we derive the sample
%  of short-period AM~CVn systems that can be detected in optical
%  and/or X-rays. 
  Using a self-consistent model for the star formation history and
  radial distribution of stars in the Galaxy plus simple models for
  the emission of optical and X-ray radiation from the AM~CVn systems
  and interstellar absorption, we derive the sample of short-period AM
  CVn systems that can be detected in the optical and/or X-ray bands.
  At the shortest periods the detectable systems are all X-ray
  sources, some with periods as short as three minutes. At periods
  above 10 minutes most detectable systems are \rev{optical sources}.
  About one third of the X-ray sources are also detectable in the
  optical band. We also calculate the gravitational wave signal of the
  short-period AM~CVn systems. We find that \rev{potentially} several
  thousand AM~CVn systems can be resolved by the gravitational wave
  detector \textit{LISA}, comparable to the \rev{expected} number of
  detached double white dwarfs that can be resolved.  We estimate that
  several hundreds of the AM~CVn systems resolvable by \textit{LISA},
  are also detectable in the optical and/or X-ray bands.
\end{abstract}


\section{Introduction}

AM~CVn systems are binaries, with periods less than 1 hour, in which a
white dwarf accretes matter from a low-mass, helium-rich, object. The
mass transfer is driven by gravitational wave radiation
\citep[e.g.][]{pac67}. They are observed at optical wavelengths as
faint blue and variable objects.  For a reviews of observational work
on AM~CVn systems, see \citet{war95,sol03}. A discussion of the
formation and modelling of the population can be found in
\citet{ty96}, \citet{npv+00}, hereafter \citetalias{npv+00} and
\citet*{phr01}.  \citet{hb00} and \citet*{nyp01} discussed the
contribution of the AM~CVn population to the Galactic unresolved
gravitational wave background.

The number of known or suspected systems is rapidly increasing: six
new (candidate) AM~CVn systems have been found in the last five years:
CE 315, with a period of 65.1 minutes \citep{rrg+01}; KL Dra, with a
period of 25 minutes \citep{jgc+98,wcg+02}; ``2003aw'', with a
possible period of 34 minutes \cite{ww03a}; RX~J1914.4$+$2456 (V407~Vul),
with a period of 9.5 minutes \citep{chm+98,rcw+00,rwc02,ms02}, KUV
01584$-$0939 (ES~Cet) with a period of 10.3 minutes \citep{ww02} and
RX~J0806.3$+$1527 with a period of 5.3 minutes \citep{ihc+02,rhc02}.
RX~J1914.4$+$2456 and RX~J0806.3$+$1527 were discovered as
\textit{ROSAT} X-ray sources \citep{mhg+96,ipc+99}. For these two
systems the interpretation as AM CVn systems has been questioned
\citep{wcr02,nhw02}, but not for the other four new systems.

The last three of the new systems mentioned above are particularly
interesting as theoretical models of the evolution of AM~CVn systems
and their current Galactic population
\citep[e.g.][\citetalias{npv+00}]{ty96} show that the expected number
of systems at periods shorter that $\sim$10 minutes is small, and an
optically selected sample will be dominated by longer period systems.
However, at these short periods the mass transfer rate is as high as
$10^{-6} - 10^{-8}$ \msun yr$^{-1}$, \rev{and therefore systems might
  be discovered as X-ray, rather than optical sources,}
%so the emission from these
%systems is likely mostly in X-rays, 
as is the case for the two \textit{ROSAT} sources. At these short
periods the systems are also promising sources for low-frequency
gravitational wave (GW) detectors \citep[e.g.][]{ihc+02,ms02}.

In this paper we determine the range of orbital periods in which AM
CVn systems could be discovered as X-ray sources, using simple,
average temperature and blackbody emission models and taking into
account interstellar absorption. We estimate the number of such
systems in the Galaxy and their brightness in the optical band. We
compute the number of short-period ($P <$ 1500~s) AM~CVn systems that
could be detected by the planned low-frequency gravitational wave
radiation (GWR) detector in space, \textit{LISA}, and focus on the
subset of these systems that are expected to have optical and/or X-ray
counterparts, as combined observations will lead to a better
understanding of the physical state, formation and population of AM
CVn systems. We will use the term ``AM~CVn system'' throughout the
paper for all double-degenerate interacting binaries with helium
secondaries, irrespective of their optical properties.

In Sect.~\ref{AMCVnpop} we review our Galactic model of AM~CVn
systems, discuss the changes we made with respect to
\citetalias{npv+00} and the uncertainties in this model. The emission
of optical, X-ray and gravitational wave signals from AM~CVn systems
is discussed in Sect.~\ref{emission}. The selection effects and the
resulting expected population of sources detectable in the optical and
X-ray bands, and as GWR sources are presented in Sect.~\ref{results}.
In Sect.~\ref{counterparts} we consider the \textit{LISA} sources that
have optical and/or X-ray counterparts. Discussion
(Sect.~\ref{discussion}) and conclusions (Sect.~\ref{conclusion})
follow.

\section{A model for the current Galactic population of AM CV\lowercase{n} systems}\label{AMCVnpop}

\subsection{Formation of AM~CVn systems}\label{popintro}

For a detailed discussion of the formation of AM~CVn systems we refer
to \citetalias{npv+00}, and references therein, so here we only give a
brief summary. There are two main routes for the formation of AM~CVn
systems: (i) via a phase in which a double white dwarf loses angular
momentum due to gravitational wave radiation and evolves to shorter
and shorter periods to start mass transfer at periods of a few minutes
and (ii) via a phase in which a non-degenerate helium star transfers
matter to a white dwarf accretor.

Recently \citet{phr01} proposed a third scenario to form AM~CVn
binaries, \rev{similar to the usual scenario for the formation of
  cataclysmic variables (CVs),} via a phase of mass transfer in the
course of which a low-mass star with initially almost exhausted
hydrogen core or a small helium core becomes a helium white
dwarf\footnote{Such a scenario has been proposed before
  \citep[e.g.][]{tfe+87} as a formation scenario for ultra-compact
  X-ray binaries}. We found that \rev{with the assumptions used in
  this paper (see below)} this scenario is relatively unimportant for
the short-period systems discussed here \rev{(giving less than two per
  cent of the systems)}, so for the remainder of the paper we leave it
out of our model, but briefly come back to it in the discussion (Sect.
\ref{discussion}).

One of the main conclusions of \citetalias{npv+00} was that the
uncertainties in the models for the formation of AM~CVn systems
resulted in widely varying numbers of expected systems in the Galaxy.
The most optimistic and most pessimistic cases of \citetalias{npv+00}
gave roughly 80 and 18 million systems in the Galaxy at present. The
uncertainty is particularly severe for the shortest period AM~CVn
systems, which probably have to form through the double white dwarf
scenario.  In \citetalias{npv+00} we showed that at the start of the
mass transfer between double white dwarfs there is a phase of ``direct
impact accretion'', in which the gas stream flowing from the donor to
the accretor hits the accretors surface directly without forming an
accretion disc \citep[see also][]{web84}. In this phase the stability
of the mass transfer depends crucially on the tidal (or other)
coupling between the spin of the accretor and its companion
\citepalias[][ see for a detailed discussion of the stability
\citealt*{mns02}]{npv+00}.

The formation of systems formed via the helium star scenario is also
uncertain because of the possibility of so called edge-lit-detonation
(ELD) of the accreting white dwarf before the system reaches the AM
CVn phase \citepalias[see][]{npv+00}. In the current study we use the
most optimistic model \citepalias[Model II of ][]{npv+00}, in which we
assume efficient coupling in the direct impact phase, and that
accretion of at least 0.3 \msun of He is necessary for ELD. But we
note that recent calculations, including rotation, of helium accreting
white dwarfs suggest that ELD might not happen at all \citep{yn03}.
This would increase the formation rate of AM~CVn systems via the
  helium star scenario in our most optimistic model by about 15 per
  cent. For the short period systems discussed here the total
uncertainty is probably of the order of a factor 10 at least.

We note that we do not take into account the unstable helium burning
on the surface of carbon-oxygen white dwarfs accreting helium at high
rates and the possible associated mass loss \citep[see, e.g.
][]{kh99}.

\subsection{Galactic model, star formation history and stellar
  evolution}\label{galmodel}

\begin{figure}
  \resizebox{\columnwidth}{!}{\includegraphics[angle=-90,clip]{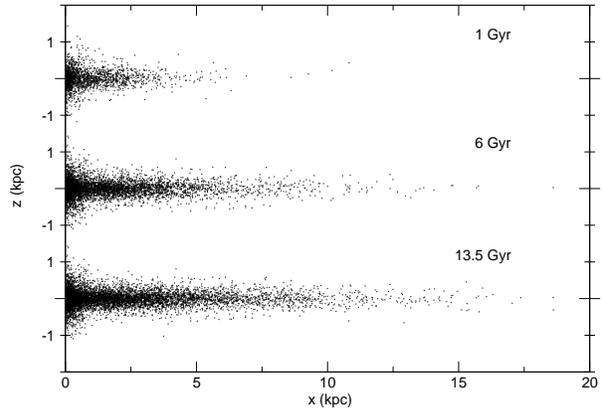}}
\caption[]{The distribution of stars in the $x - z$ plane of the
  Galaxy for our implementation of the \citet{bp99} model at Galactic
  ages of 1, 6 and 13.5 Gyr, using a number of tracer particles
  proportional to the total mass in stars ($M_*$) in the Galaxy as
  indicated in the plot.  The change in the radial extent of the star
  forming region in time is clearly visible. For the total SFR as
  function of time, see Fig.~\ref{fig:br_AMCVn}.}
\label{fig:x_z_BP}
\end{figure}

\begin{figure*}
\resizebox{\textwidth}{!}{\includegraphics[angle=-90,clip]{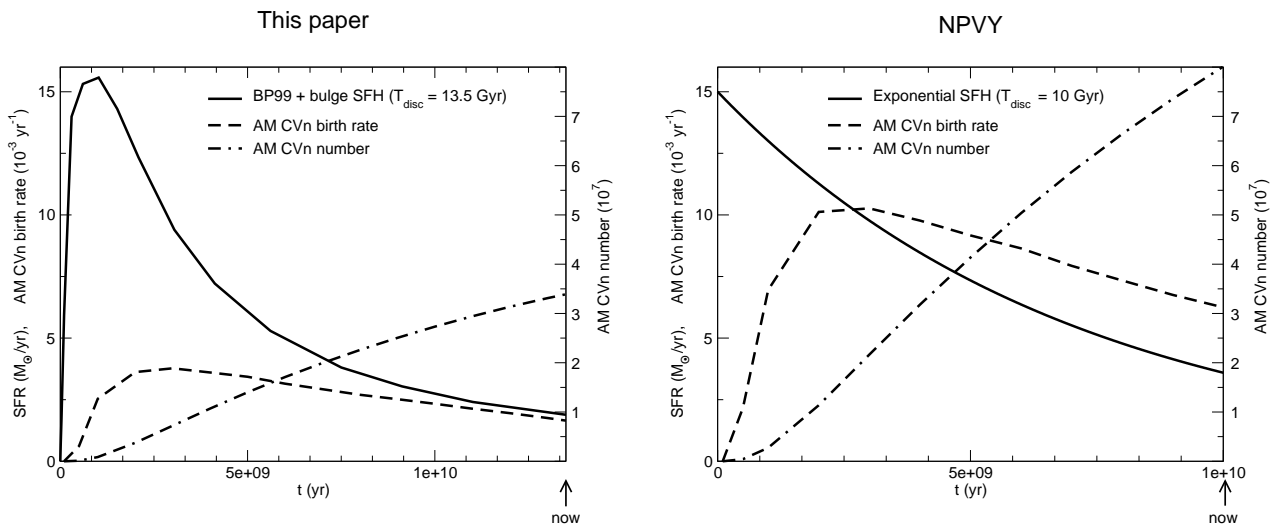}}
\caption[]{
  Illustration of the difference in Galactic model between the current
  work (left panel) and \citetalias{npv+00} (right panel).  Shown are
  the SFR (solid lines), the birth rate of AM~CVn systems (dashed
  line, left ordinate), and their total number (dot-dashed line, right
  ordinate) as function of time in the Galaxy. The differences in the
  birth rate and number of AM~CVn systems reflect the differences in
  star formation history and the treatment of stellar evolution (see
  Sect.~\ref{totpop}).  Note that in the exponential model the
  Galactic age is 10 Gyr, while the \citet{bp99} model has an age of
  13.5 Gyr.}
\label{fig:br_AMCVn}
\end{figure*}

Instead of the simple Galactic model we used before, we implemented a
star formation rate (SFR) as function of time and position in the
Galaxy based on the model of \citet{bp99}. We obtain the SFR as
function of $R$ and $t$ by linear interpolation between values in a
table (S. Boissier, private communication), and assume that the
probability of a star being born \rev{with a radius $R$ or
  smaller} goes with the integrated SFR, i.e.
\begin{equation}
P(R, t) = \frac{\int_0^R \mbox{SFR}(R', t) \, 2 \, \pi \, R' \, {\rm d} R'}{\int_0^{R_{\rm max}} \mbox{SFR}(R',
  t) \, 2 \, \pi \, R' \, {\rm d} R'},
\end{equation}
where $R_{\rm max}$ is the maximum extent of the disc (19 kpc) and we
assume that the stars do not migrate radially in time. The age of the
Galaxy in this model is 13.5 Gyr. As the \citet{bp99} model gives the
SFR projected onto the plane of the Galaxy, we assume a $z$-distribution,
\begin{equation}\label{eq:Pz}
P(z) \propto \mbox{sech}(z/z_h)^2,
\end{equation}
as in \citet{nyp+00}, where $z_h$ = 200 pc, neglecting its age and
mass dependence. We have added a bulge to the model, simply by
doubling the SFR in the inner 3 kpc of the Galaxy compared to
\citet{bp99}. As a result the total mass in this region is 2.6
10$^{10}$ \msun at the current age of the Galaxy, consistent with
kinematic and micro-lensing results \citep[e.g.][]{kzs02}. We
distribute the stars in the bulge according to
\begin{equation}
\rho_{\rm bulge}  = \exp(-(r/0.5 \mbox{ kpc})^2),
\end{equation}
where $r = \sqrt{x^2 + y^2 + z^2}$. 

In Fig.~\ref{fig:x_z_BP} we plot the resulting distribution of stars
in the $x - z$ plane of the Galaxy, showing the $z$-distribution of
stars and the change in the radial extent of the star formation as
function of time. For the SFR integrated over $R$ as function of time,
see Fig.~\ref{fig:br_AMCVn} (left panel).

\begin{figure}

  \resizebox{\columnwidth}{!}{\includegraphics[angle=-90,clip]{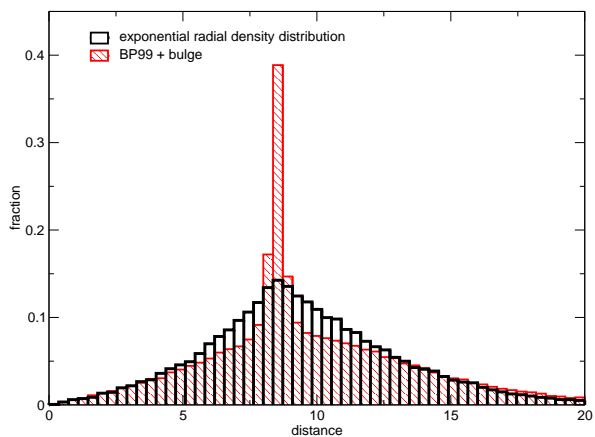}}
\caption[]{ Histogram of distances to the Earth of short period AM~CVn systems for our
  implementation of the \citet{bp99} model \rev{with added bulge} and
  for an exponential radial density distribution.}
\label{fig:compare_d}
\end{figure}

The important effect resulting from the different choice of Galactic
model is that for most of the time in the Galactic history, the SFR is
lower in the \citet{bp99} model than in the `exponential' model used
in \citetalias{npv+00} (see Fig.~\ref{fig:br_AMCVn}); in particular,
the current SFR is lower than the one we obtained previously (1.9 vs
3.6 \msun yr$^{-1}$). This has a strong effect on the current birth
rates of AM\,CVn systems, since the latter follow star formation
history with a typical delay of several Gyr \citep{ty96,npv+00}.
However, the integrated SFR remains quite similar ($8.2 \times
10^{10}$ vs. $8 \times 10^{10}$ \msun), partly due to the different
Galactic ages (13.5 vs. 10 Gyr).  Another characteristic of our new
model is that the Galactic distribution is more centrally concentrated
than with an exponential radial density distribution\rev{, mainly due
  to the inclusion of the bulge.} In Fig.~\ref{fig:compare_d} we show
the distribution of distances from the Earth of the population of
short-period AM~CVn systems in the Galaxy for the two models. The
concentration in the Galactic centre with the new model is clearly
visible. As the longer period AM~CVn systems are older, their
distribution is even more centrally concentrated

Another new element in our current study is the use of the IMF as
proposed by \citet*{ktg93}, rather than the \citet{ms79} IMF as used
in \citetalias{npv+00}. \rev{Because the \citet{ktg93} IMF is steeper
  for masses above 1 \msun, the same star formation history leads to
  fewer AM CVn systems, especially for the helium star formation
  scenario, the efficiency of which is reduced by about 25 per cent
  (see Sect.~\ref{totpop}).}

We use the population synthesis code \textsf{SeBa}, developed by
\citet[see also \citealt{nyp+00}]{pv96}. We assume a 50 per cent
binary fraction, an initial distribution of separations in binary
systems ($a$) flat in $\log a$ between contact and 10$^6$ R$_\odot$, a
flat mass ratio distribution and a thermal eccentricity
distribution.\footnote{\rev{All binaries discussed here will be
    circularized to somewhat smaller separation before the first phase
    of mass transfer and thus the shape of the initial eccentricity
    distribution hardly influences the number of AM CVn systems.}} The
current study is based on a model with 250,000 ZAMS binaries.  For the
treatment of the stellar evolution and the mass transfer we refer to
our earlier papers, but recall that we use a common envelope formalism
based on the angular momentum balance for the first episode of
unstable mass transfer in the system, as detailed in
\citet{nvy+00,nyp+00}. Two small changes that turn out to be important
are in the formation and evolution of white dwarfs. We updated our
treatment of the growth of the core in giants as described in
\citet{nyp+00}, by including a simple model of the second dredge-up to
be more in line with detailed stellar evolution models
\citep[e.g.][]{hpt00}.  The second change is in the treatment of white
dwarf cooling (see Appendix~\ref{appendix}) and white dwarf mass --
radius relation, for which we now use the fit by Eggleton, as given in
\citet{vr88}, rather than the \citet{zs69} relation, which is not
appropriate for massive white dwarfs.

The last change we made compared to \citetalias{npv+00} is in the
maximum rate at which a main sequence star can accrete before it
expands and mass will be lost from the system. We used to limit this
to accretion at the thermal time scale of the accretor. However, as is
shown e.g.  by \citet{km77}, higher accretion rates do not necessary
lead to expansion of the accretor. We therefore use the treatment of
\citet{pm94} which takes the reaction of the radius of the accreting
star upon mass gain into account.

\subsection{Total population of AM~CVn systems in the Galaxy}\label{totpop}

\begin{figure}
\resizebox{\columnwidth}{!}{\includegraphics[angle=-90]{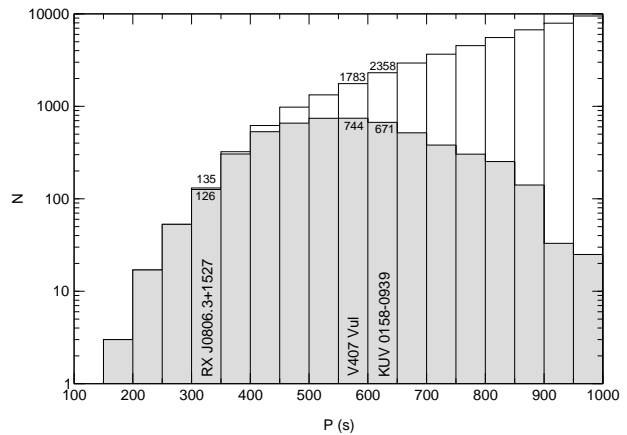}}
\caption[]{
  Histogram of periods of the shortest period AM~CVn systems. The
  shaded histogram gives the number of systems which are in the direct
  impact phase (but note the logarithmic y-scale). The observed
  systems are indicated at their suggested orbital periods.  The
  number of systems in these bins are indicated.}
\label{fig:histP}
\end{figure}

\begin{table}
\begin{center}
\begin{tabular}{lrr}
\hline
Formation channel  & birth rate & total number  \\
 & (10$^{-3}$ yr$^{-1}$) & (10$^7$) \\
\hline
double white dwarfs & 1.3 & 2.3 \\ 
helium stars & 0.27 & 1.1 \\
\hline
total & 1.6 & 3.4 \\
\hline
\end{tabular}
\end{center}
\caption[]{Present day birth rates and total number of AM~CVn systems in the
  Galaxy in our current model, for the different formation channels.}
\label{tab:br_number}
\end{table}

With the changes described in the previous section, we obtain, for the
two formation channels discussed here, the current birth rates and
total number of AM~CVn systems in the Galaxy as shown in
Table~\ref{tab:br_number} and Fig.~\ref{fig:br_AMCVn} (compare Table~1
in \citetalias{npv+00}).  The difference between the numbers presented
here and in \citetalias{npv+00} \rev{are due to the use of the new
  Galactic mode, different IMF and the changes in the treatment of
  stellar evolution described in Sect.~\ref{galmodel}. All these
  change the birth rate of AM~CVn systems at any time, e.g. at the
  current Galactic age the birth rate of the white dwarf family goes
  down by $\sim$60 per cent due to the Galactic model and is reduced
  further by another $\sim$40 per cent due to the changes in treatment
  of white dwarf formation and mass -- radius relation.  The birth
  rate of the helium star family goes down by $\sim$60 per cent due to
  the Galactic model, $\sim$25 per cent due to the IMF and $\sim$40
  per cent due to the higher accretion rate limit for main-sequence
  stars (because less systems pass through common envelopes and become
  close enough to produce AM~CVn systems). Partly compensated by the
  change in the age of the Galaxy this results in 55 and 65 per cent
  fewer systems present in the Galaxy for the two families,
  respectively.}

In Fig.~\ref{fig:histP} we show a histogram of the shortest period
systems as function of orbital period, indicating the systems in the
direct impact phase and the periods of the observed (candidate) AM~CVn
systems.

\section{Modelling the optical, X-ray and GW emission of
  AM CV\lowercase{n} systems}\label{emission}

In order to be able to compare our \rev{results} with (future)
observations we have to \rev{simulate} the emission produced by our
model systems. We determine the temperature and luminosity of the
different emission sources and determine how much energy is emitted in
a certain band, assuming the emitted spectrum is a blackbody. For the
present study we only discuss emission in $V$-band and the
\textit{ROSAT} 0.1--2.4 keV X-ray band plus gravitational wave
radiation. \rev{We chose the \textit{ROSAT} band, because the two
  candidate systems were found as \textit{ROSAT} sources and because
  we can compare our results to the \textit{ROSAT} all-sky survey.
  However, in Sect.~\ref{discussion} we briefly discuss the expected
  results for the \textit{Chandra} and \textit{XMM}.}

There are four main emission sites: the accretion disc and boundary
layer between disc and accreting white dwarf, the impact spot in the
case of direct impact accretion, the accreting star and the donor
star.

\subsection{Optical emission}\label{optical}

\begin{figure}
  \resizebox{\columnwidth}{!}{\includegraphics[]{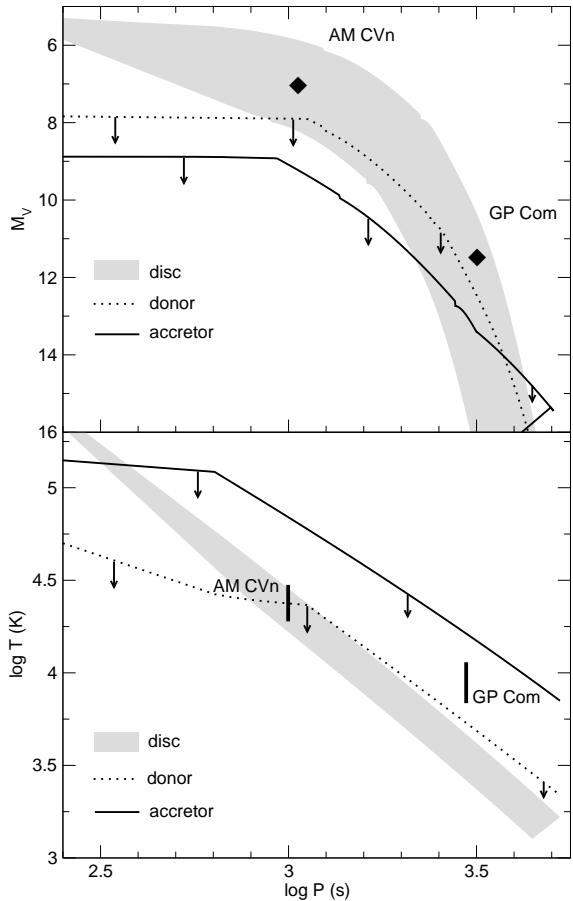}}
\caption[]{Absolute magnitudes and temperatures of the different
  optical emission components as function of orbital period. The
  limits of the regions are defined by systems with initial masses of
  0.15 + 0.25 \msun and 0.3 + 1.0 \msun. For the accretor and donor we
  give the upper limits on the magnitude and temperature for the
  extreme case that the donor/accretor white dwarf was young and hot
  at the start of the AM~CVn phase, i.e. at the top of the cooling
  curve (see Appendix~\ref{appendix}).}
\label{fig:Mv_T_opt}
\end{figure}

\begin{description}
\item[\textbf{The disc.}] 
We assume that half of the accretion luminosity is
  radiated by the disc (the other half in the boundary layer), i.e.
\begin{equation}\label{eq:L_acc}
L_{\rm disc} = \mbox{$\frac{1}{2}$} G \, M \, \dot{m}
\left(\frac{1}{R} - \frac{1}{R_{\rm L1}}\right) \mbox{erg s}^{-1},
\end{equation}
where $M$ and $R$ are the mass and radius of the accretor and $R_{\rm
  L1}$ is the distance of the first Lagrangian point to the centre of
the accretor and $\dot{m}$ is the mass transfer rate.  Because the two
white dwarfs are so close together the usual assumption that the
matter falls in from infinity is not valid \citep{hw99}.  We model the
optical emission from the disc like in \citetalias{npv+00}, according
to \citet{wad84}, as a single temperature disc that extends to 70 per
cent of the Roche lobe of the accretor, radiates as a blackbody and
use the bolometric correction appropriate for its temperature.

\item[\textbf{The donor.}] We model the emission from the donor as the
  cooling luminosity of the white dwarf,
%plus  the possible heating of the white dwarf due to the proximity of the
%  accretor, according to $T_{\rm donor} \approx (R/a)^{1/2} T_{\rm
%    accretor}$.  
%  We model the cooling of white dwarfs 
  using \rev{approximations} to the models of \citet{han99}, see the
  Appendix~\ref{appendix}.

  \item[\textbf{The accretor.}] We model the emission from the accretor
  as being the unperturbed cooling luminosity of the white
  dwarf. 
\end{description}

In our simple model we do not take into account such effects as
irradiation of the donor by the impact spot or disc, heating of the
accretor by radiation from the disc/boundary layer
\citep[e.g.][]{pri88}, compressional heating \citep[e.g.][]{tb02} \rev{or
tidal heating of the two white dwarfs prior to contact \citep{itf98}}.

In Fig.~\ref{fig:Mv_T_opt} we show the range of absolute
$V$-magnitudes (M$_{\rm V}$) and temperatures of the different optical
components described above, for systems forming from double white
dwarfs. The top panel shows the range of absolute magnitudes of the
discs as function of period (in grey). The two symbols show the
observed absolute magnitudes for AM~CVn and GP~Com [using distances of
235 pc for AM~CVn (C. Dahn\footnote{On behalf of the USNO CCD Parallax
  Team.}, private communication) and 70 pc for GP~Com \citep{tho03}],
while the bars in the bottom panel show the range of effective
temperatures for these systems.  The upper and lower limit of the
shaded regions are for \rev{two combinations of masses that bracket
  the range of masses of the majority of the systems: 0.3 + 1.0 and
  0.15 + 0.25 \msun respectively}.

\rev{Because for the accretor and donor we assume that their luminosity is
the cooling luminosity, their magnitudes depend on their ages, rather
than on the orbital period. However, during the AM CVn phase there is
a direct relation between time and period (e.g. figure~4 in
\citetalias{npv+00}) so for given cooling luminosity at the start of
the mass transfer we can plot the magnitude of the cooling white dwarf
as function of orbital period. In Fig.~\ref{fig:Mv_T_opt} we plot that
curve for the extreme case in which the cooling luminosity at the
starts of the mass transfer is that of a newly born, very hot and
luminous white dwarf.}  In reality the donor and accretor are typically
of the order of a Gyr old, so they evolve along a track that lies
below these curves.  Towards the longer periods, the age of the system
is a few Gyr anyway, \rev{so the curves for all initial luminosities
converge}.

%For the accretor and donor we give the upper limits on the magnitude
%and temperature as the solid and dotted lines, which are defined
%by the curves along which they evolve if either of them is very young
%and hot at the start of the AM~CVn phase.  In reality the donor and
%accretor are typically of the order of a Gyr old, so they evolve along
%a track that lies below these curves. Towards the longer periods, the
%age of the system is a few Gyr anyway, so the donor and accretor
%luminosities become close to the upper limits.
%%most extreme case is assumed, i.e. that both white dwarfs are formed
%%just before the mass transfer starts and thus are still very hot.
%%Depending on the cooling time before the stars get into contact the
%%stars will have cooled down to be dimmer and cooler at given period.
%%The peculiar shapes of the curves for the donor and accretor are an
%%artefact of the extremely simple way of modelling the white dwarf
%%cooling (see Fig.~\ref{fig:wd_cooling}). 

\rev{The few observed systems fall in the grey area in
  Fig.~\ref{fig:Mv_T_opt} indicating that our simple model is at least
  able to understand these systems satisfactorily except the
  temperature of GP~Com\footnote{Indeed the continuum emission of GP
    Com originates at the reheated accreting white dwarf, rather than
    in the disc (see Bildsten et al., in preparation).}  The same
  model can be adopted for the systems that form via the helium-star
  donors. In these cases, however, the curves shift to longer periods
  (see \citetalias{npv+00}).}

%The observations show that our simple model gives reasonable estimates
%of the optical emission.  For the systems forming from helium star
%donors, we use the same model, which would give similar ranges as
%shown in Fig.~\ref{fig:Mv_T_opt}, but at slightly longer periods
%(compare Fig.~5 of \citetalias{npv+00}).

\subsection{X-ray emission}\label{X-ray}

\begin{figure}
\resizebox{\columnwidth}{!}{\includegraphics[]{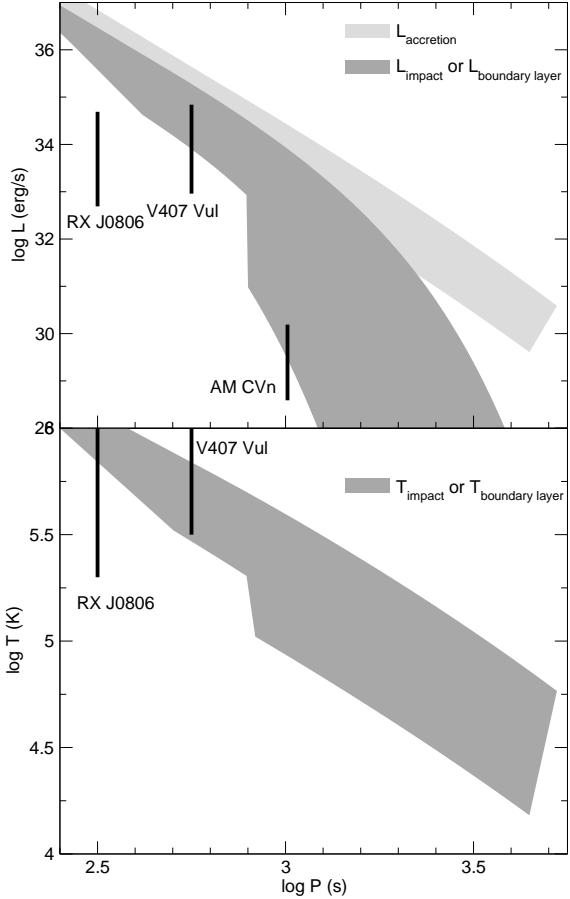}}
\caption[]{\label{fig:L_T_X}
  Total accretion luminosity and luminosity and temperatures of the
  X-ray emission components (impact spot or boundary layer) as
  function of orbital period. The limits of the regions are again
  defined by systems with initial masses of 0.15 + 0.25 \msun and 0.3
  + 1.0 \msun.}
\end{figure}

In modelling the X-ray emission of AM~CVn systems we distinguish two
cases: with and without disc.

If there is no disc, the gas stream hits the companion directly and a
small area of the surface will be heated. We assume that the full
 accretion luminosity is radiated as a blackbody with a
temperature given by
\begin{equation}
\left(\frac{T_{\rm BB}}{T_{\odot}}\right)^4 = \frac{1}{f} R^{-2} L_{\rm acc},
% = \frac{G \, M \, \dot{m}}{f\, R^2}  \left(\frac{1}{R} - \frac{1}{R_{\rm L1}}\right),
\end{equation}
where $L_{\rm acc}$ and $R$ are in solar units \rev{and $L_{\rm
    acc}$ is defined by Eq.~(\ref{eq:L_acc})}. The fraction $f$ of the
surface that is radiating depends on the details of the accretion, but
will in general be small. We used $f = 0.001$, which is consistent
with the observed X-ray emission of V407~Vul and simple arguments
regarding the impact stream \citep{ms02}.

If there is an accretion disc, the X-ray emission can come from the
boundary layer between the inner disc and the accretor, where we
assume half of the accretion luminosity is radiated, as suggested e.g.
by \citet{pri77}, or it might have a non-thermal origin. The average
temperature of the disc \citep{wad84}, which we use to estimate the
optical emission is never high enough to produce X-rays
\citep{bep+74}. Even though from observations of cataclysmic variables
it is clear that the real X-ray emission mechanism is much more
complicated \citep[e.g.][]{hd91,mwp+91,vv94,vbv96}, we model the X-ray
emission of systems with a disc for the moment as coming from the
boundary layer, with temperature \citep{pri77}
\begin{equation}
T_{\rm BL} = 5 \times 10^{5}  
            \left(\frac{\dot{m}}{10^{18}
                \mbox{ g/s}}\right)^\frac{2}{9} 
            \left(\frac{M}{M_\odot}\right)^\frac{1}{3}
            \left(\frac{R}{5 \times 10^8 \mbox{ cm}}\right)^{-\frac{7}{9}}
            {\rm K}.
\end{equation}

In Fig.~\ref{fig:L_T_X}, top panel, we show how the luminosity in
X-rays compares to the total accretion luminosity. For the X-ray
luminosity we consider here the emission in the \textit{ROSAT} band,
from 0.1 -- 2.4 keV. The temperature of the impact spot or boundary
layer is shown in the bottom panel. The sharp drop in the X-ray
luminosity and temperature at the bottom of the darker region \rev{at
  $\log P \approx 2.9$} is caused by the change from direct impact to
  disc accretion in the low-mass system. We assume that the transition
  from the impact temperature to the boundary layer temperature is
  instantaneous. This transition for the more massive system occurs
  just left off the plot, where the evolution is so fast that hardly
  any systems are expected \rev{to be observed}.
  
  At periods larger than 1000 -- 2000 s the mass transfer rates drop
  to below 10$^{-10}$ \msun yr$^{-1}$, the temperature drops below
  10$^5$ K and hardly any X-ray emission is produced in the boundary
  layer. The inferred temperatures and X-ray luminosities of RX~J0806
  and V407~Vul are plotted as the bars in the figures. For RX~J0806,
  we use the X-ray flux ($4 \times 10^{-10}$ erg cm$^{-2}$ s$^{-1}$),
  distance (0.1 -- 1 kpc) and blackbody temperature ($6^{+13}_{-4} \times
  10^5$ K) as given by \citet{ihc+02}. For V407~Vul blackbody
  temperatures of 43 and 56 eV are derived by \citet{mhg+96} and
  \citet{rwc02} respectively, while the latter authors derive an X-ray
  luminosity of $1.2^{+4.1}_{-0.3} \times 10^{33}$ erg s$^{-1}$, for a
  distance of 100 pc and we plot these limits for a distance range 100
  -- 400 pc \citep{rcw+00}.  The X-ray luminosity of AM~CVn is taken
  from \citet{ull95a}. Except for the luminosity of RX~J0806, our very
  simple model agrees well with the observations.

\subsection{Gravitational wave emission}

We assume that AM~CVn systems emit GWR as two point sources, i.e.
neglecting any effect of the deformation of the donor, which fills its
Roche lobe and the mass in the accretion disc, if present
\citep[see][for a discussion of the small effect of deformation for
cataclysmic variables]{ruy01}. The amplitude of the GWR signal, as can
be detected by \textit{LISA}, is given by \citep[e.g.][]{eis87}
\begin{equation}\label{eq:h}
h = 0.5 \times 10^{-21}  \left( \frac{\mathcal{M}}{\msun}
\right)^{5/3} \! \! \! \left( \frac{P_{\rm orb}}{\rm 1\, {\rm hr}} \right)^{-2/3}
\! \! \!\left( \frac{d}{\rm 1 {\rm kpc}} \right)^{-1},
\end{equation}
where $\mathcal{M} = (M \, m)^{3/5} / (M + m)^{1/5}$ is the chirp
mass, $P_{\rm orb}$ is the orbital period and $d$ is the distance to
the system.

%\begin{itemize}
%\item direct impact
%\item accretion luminosity L1 deep in potential well
%\item X-ray emitting region: fraction of WD surface
%\item What happens when disc forms: average disc temperature, but hot
%  inner part ==> X-rays?
%\end{itemize}

%\begin{figure*}
%%\resizebox{1.6\columnwidth}{!}{\includegraphics[angle=-90]{lb_new.ps}}
%\resizebox{1.6\columnwidth}{!}{\includegraphics[angle=-90]{ultracompact_lb.ps}}
%\caption[]{Distribution of ``detectable'' short period AM~CVn
%  systems on the sky in Galactic coordinates. Symbols as in
%  Fig.~\ref{fig:Pd}.}
%\label{fig:lb}
%\end{figure*}

\section{Results I: the detectable population of short-period AM CV\lowercase{n} systems}\label{results}

\subsection{Selection effects}\label{selection}

To compare our models with the number of observed systems, we have to
decide which of the model systems would be detectable.

For the optical emission we focus on model systems that are brighter
than $V = 20$ mag, which is similar to the optical magnitude of the
observed short period AM~CVn candidates. To take the interstellar
reddening into account we use the \citet{san72} model
\begin{eqnarray}
A_{\rm V} (\infty) & = & 0.165 \, \frac{(\tan(50\degr) - \tan(b))}{\sin(b)}  \quad \mbox{ for } b < 50\degr  \nonumber \\
  & = & 0  \qquad \qquad \qquad \qquad \qquad \qquad \mbox{ for } b \ge 50\degr,
\end{eqnarray}
and $A_{\rm V} (d)$ follows from our Galactic model
[Eq.~(\ref{eq:Pz})]
\begin{equation}
A_{\rm V} (d) = A_{\rm V} (\infty) \tanh\left(\frac{d \sin(b)}{z_h}\right).
\end{equation}

For the sources detectable in X-rays we select the systems with an
X-ray flux in the 0.1 -- 2.4 keV (\textit{ROSAT}) band higher than
$10^{-13}$~erg~s$^{-1}$~cm$^{-2}$. To estimate this flux we use the
intrinsic flux in this band (Sect.~\ref{X-ray}), the distance and an
estimate of the Galactic hydrogen absorption \citep{mm83}. We modelled
the hydrogen column density after \citet{ps95} as
\begin{equation}
N_{\rm H} = 0.179 \, A_{\rm V} (d) \, 10^{22} \quad {\rm cm}^{-2}.
\end{equation}
%with $A_{\rm V} (d)$ given above.

\subsection{Systems detectable in the optical and X-ray bands}

\begin{figure}
%\resizebox{\columnwidth}{!}{\includegraphics[]{Pd_new.ps}}
\resizebox{\columnwidth}{!}{\includegraphics[]{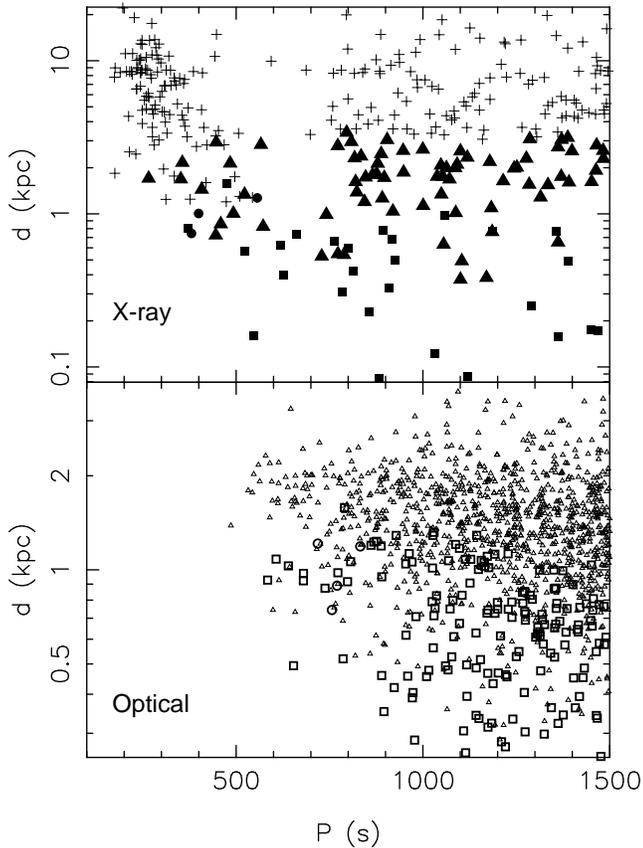}}
\caption[]{
  Period vs distance distribution of systems detectable in the X-ray
  and/or optical bands. \textbf{Top panel}: Systems detectable only in
  X-rays (pluses), and systems that are detectable in the optical and
  X-ray band (filled circles: direct impact systems with a bright
  donor, filled triangles: systems with an optically bright disc,
  filled squares: systems with bright disc + donor). \textbf{Bottom
    panel}: Systems only detectable in the optical band (circles:
  direct impact systems with a bright donor, triangles: bright discs,
  squares: bright disc + donor).  }
\label{fig:Pd}
\end{figure}

In Fig.~\ref{fig:Pd} we show the sample of the systems that would be
detectable in the optical and/or X-ray bands, taking into account the
selection effects described above. The relevant numbers are given
below in the text, between square brackets (see also
Table~\ref{tab:resolved}). (We should recall that the results we
discuss represent one possible random realization of the model for
current Galactic population of AM~CVn systems, so all numbers given
are subject to Poisson noise). We limited ourselves to the short
period systems, with $P < 1500$ s.  At longer periods the optical
emission from the disc will completely dominate and the detectable
sample has been discussed in \citetalias{npv+00}.  Fig.~\ref{fig:Pd},
top panel, shows the systems detectable in X-rays.  Pluses are [220]
systems only detectable in X-rays, while the filled symbols mark
systems also detectable in the $V$-band.

At the shortest periods a significant fraction of the small number of
systems that exist (see Fig.~\ref{fig:histP}) produces enough X-rays
that they are detectable even though the sources are close to the
Galactic centre (Fig.~\ref{fig:Pd}). At longer periods the X-rays get
weaker (and softer) so only the systems close to the Earth can be
detected.  Several of these systems are also detectable in the optical
band (filled symbols). The [3] filled circles show the systems in the
direct impact phase, where the optical emission comes solely from the
donor as the impact spot is too hot to emit in the $V$-band.  Systems
with optical emission from a disc are shown as triangles [75].  There
are some [28] systems that are close enough to the Earth that the
donor stars can be seen as well as the discs (filled squares).  Above
600 s the systems descending from helium star donors appear and have
high enough mass transfer rate to be X-ray sources, the closer ones
also being visible in the $V$-band (filled triangles/squares, as these
systems always have a disc).

The X-ray sources at larger distances (pluses) are almost all in the
bulge and thus are concentrated in a small area on the sky. The closer
systems, which often are also detectable in the $V$-band are more
evenly distributed over the sky.

The bottom panel of Fig.~\ref{fig:Pd} shows the 1230 `classical'
AM~CVn systems, detectable only by the optical emission, for most
systems only from their accretion disc (open triangles [1057]). Of
this population the ones closest to the Earth in some cases [169] also
have a visible donor (open squares). A few [4] direct impact systems
only have a visible donor, but are not X-ray sources (open circles).
\rev{The majority of the optically detectable systems represented in
  the lower panel of Fig.~\ref{fig:Pd} (with orbital periods between
  1000 and 1500 s) are expected to show outbursts due to the
  viscous-thermal disc instability \citep{to97} which could enhance the
  chance of their discovery.}

Comparing these results with the tentative classification of V407~Vul
and RX~J0806 as direct impact systems, detectable in X-rays and (just)
in the optical band, shows that a few of such systems (filled circles)
could be expected in the Galaxy, given all the uncertainties in
\rev{population synthesis model}, the emission model and the selection
effects.  Furthermore it shows that many more short period AM~CVn
systems might be detectable only as X-ray sources. Some have periods
as short as three minutes. About one third of the X-ray sources is
also detectable in the optical band.

%Below periods of 1000 s the
%number of these is almost comparable to the number of ``classical''
%AM~CVn systems that can be detected due to the optical emission from
%an accretion disc. Only ten per cent of the total number of detectable
%systems is detectable both in the optical and X-ray bands.

%\begin{itemize}
%\item Fraction of BB in X-ray (ROSAT) band 0.2 - 2.4 keV
%\item X-ray absorption in the Galaxy: $N_H$ as function $A_V$
%  \citep{ps95} and model for $A_V$ from \citet{san72}. Average
%  correction as function $l$ after \citet{sfd98}?
%\item Optical emission from systems: contribution from donor and
%  accreting white dwarf.
%\item To be incorporated: compressional heating of accretor, heating
%  and irradiation of donor
%\end{itemize}

%\begin{figure}
%\resizebox{\columnwidth}{!}{\includegraphics[angle=-90,clip]{PMdot_Xray.ps}}
%\caption[]{Current Galactic population of AM~CVn systems, weighted by
%  the luminosity above 0.2 keV. The suggested periods of the observed
%  (X-ray selected) systems are shown by the dotted lines. }
%\label{fig:PMdot}
%\end{figure}

\subsection{Gravitational waves from short period AM~CVn systems}\label{GWR}

\begin{figure*}
\resizebox{2\columnwidth}{!}{\includegraphics[angle=-90,clip]{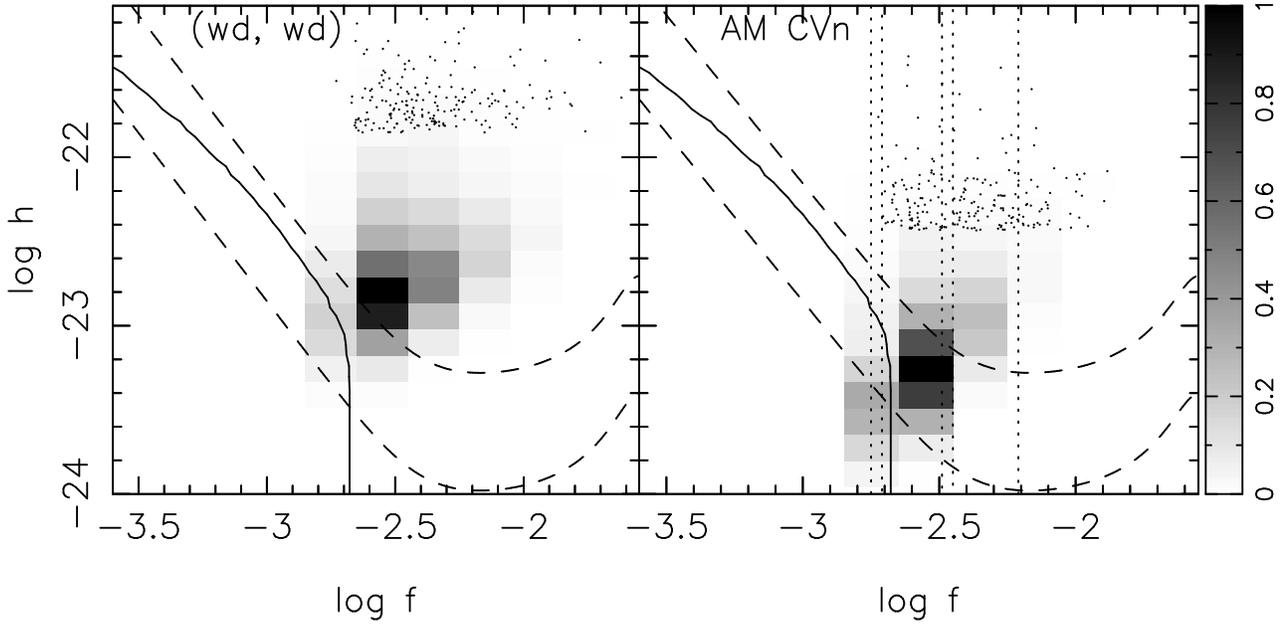}}
\caption[]{
  Gravitational wave amplitude $h$ as function of the frequency of the
  wave for the resolved double white dwarfs (left panel, 10697
  objects) and AM~CVn systems (right panel, 11000 objects.). \rev{The
    grey shades give the density distribution of the resolved systems
    normalised to the maximum density in each panel} (1548 and 1823
  for the double white dwarf and AM~CVn panel respectively). The 200
  strongest sources in each sample are shown as the dots to enhance
  their visibility. In the AM~CVn panel the periods of the observed
  short period systems are indicated by the vertical dotted lines.
  Solid line shows the average background noise due to detached white
  dwarfs as in \citet{nyp01}. The \textit{LISA} sensitivities for a
  integration time of one year and a signal-to-noise ratio of 1 and 5
  are indicated by the dashed lines \citep{lhh00}.}
\label{fig:fh}
\end{figure*}

In \citet{nyp01} we presented a study of the gravitational wave signal
from Galactic binaries \rev{in which both stars are compact objects}.
In that study we estimated the contribution of AM~CVn systems to the
double white dwarf noise background and found that AM~CVn systems
hardly contribute, even though at frequencies between 0.3 and 1 mHz
(periods between 33 and 105 minutes) they outnumber the detached
double white dwarfs. This is because at these frequencies their chirp
mass has fallen far below that of a typical detached system \citep[see
also][]{hb00}. We therefore did not consider them anymore. However at
shorter periods, where the number of AM~CVn systems is much smaller
their chirp mass is similar to that of the detached systems from which
they originate.

In Fig.~\ref{fig:fh} we show the Galactic population of detached
double white dwarfs and AM~CVn systems that can be resolved by
\textit{LISA}.  \emph{There are $\sim$11\,000 resolved double white
  dwarfs as well as $\sim$11\,000 resolved AM~CVn systems.}  Their
distribution is shown by the grey shading. The 200 strongest sources
of each type are shown as dots to enhance their visibility.
Figure~\ref{fig:fh} clearly shows that AM~CVn systems may be important
sources for \textit{LISA}. Compared to our previous study the
`confusion limit', i.e. the frequency at which the unresolved
background disappears (see \citet{nyp01} for details) moves to
slightly higher frequency because in our current model there are 213
million detached close double white dwarfs, instead of 110 million
before, \rev{due to the use of different IMF, Galactic model and the
  changes in stellar evolution}.

The crucial difference between AM~CVn systems and detached double
white dwarfs in the short period range is the sign of their period
derivatives. The periods of AM~CVn systems increase due to mass
transfer, while the periods of detached systems decrease due to GW
losses\footnote{Note that systems formed from helium stars and evolved
  secondaries in CVs have decreasing periods before they reach their
  period minimum \citep[e.g.][]{tf89,phr01}. However, in this phase
  they are (much) brighter than after the period minimum and are not
  consistent with being classified as ``classical'' AM~CVn
  systems.\label{foot}}.  \rev{Assuming conservative mass
  transfer,} the change of the GW frequency ($f = 2/P$) of AM~CVn
systems formed from double white dwarfs can be derived from
\begin{equation}
\frac{\dot{f}}{f} = - \frac{\dot{P}}{P} = -\frac{3}{2} \, \frac{\dot{a}}{a} = 
-3 \, \frac{\dot{J}_{\rm GWR}}{J_{\rm orb}} - 3 (1 - q) \frac{\dot{m}}{m},
\end{equation}
where $m$ is the mass of the donor star and $q = m/M$. We can use the
equilibrium mass transfer rate as derived in \citet{mns02} to get
\begin{equation}\label{eq:fdot_AMCVn}
\dot{f} = - 5.8 \times 10^{-7} f^{11/3} \left( \frac{\mathcal{M}}{\msun}
\right)^{5/3}
\left(\frac{\zeta_{r_L} - \zeta_2}{2 + \, \zeta_2-\zeta_{r_L} \,- 2
    q}\right) \mbox{Hz s$^{-1}$},
\end{equation}
where $\zeta_2$ is the logarithmic derivative of the secondary radius
with respect to mass \rev{and $\zeta_{r_L} = {\rm d} \ln (R_L/a) /
  {\rm d} \ln m \approx 1/3$} \citep[see][]{mns02}.  We wrote the
equation in the same form as the corresponding equation for frequency
change in detached systems \citep[e.g. Eq.~(6) in][]{nyp01}, with the
(positive) term in parentheses giving the change with respect to the
detached case. This term is of order unity at the onset of the mass
transfer (i.e. at highest frequency, above $\sim$5 mHz), decreasing to
about 0.1 at 2 mHz.

\textit{LISA} will be able to measure the frequency change of a number
of the resolved binaries. In \citet{nyp01} we estimated the number of
systems for which this can be done by requiring that in the course of
the experiment the frequency would change by one frequency bin.
However, it really is the phase shift resulting from a frequency
change that will be measured \citep[e.g.][]{wh98}, so the requirement
has to be that over the course of the experiment the phase shifts by a
certain amount, compared to a system without frequency change. Writing
the phase as an expansion in $f, \dot{f}$ and $\ddot{f}$, this means
$\dot{f}$ can be measured if
\begin{equation}
\Delta \Phi = \dot{f} \, \frac{T^2}{2} > \Delta \Phi_{\rm min},
\end{equation}
where $\Delta \Phi_{\rm min}$ is the minimum phase shift that can be
measured \rev{and $T$ is the duration of the experiment}.
Analogously $\ddot{f}$ can be measured if
\begin{equation}
\Delta \Phi = \ddot{f} \, \frac{T^3}{3} > \Delta \Phi_{\rm min}.
\end{equation}
Measuring the frequency change provides important information, as the
chirp mass can be measured directly for the detached systems. Combined
with the wave amplitude $h$ [Eq.~(\ref{eq:h})], this yields the
distance. For AM~CVn systems this argument does not hold, as $\dot{f}$
is determined by the mass transfer [Eq.~(\ref{eq:fdot_AMCVn})].
However, for an assumed mass--radius relation of the donor star,
both the model mass transfer rate and the mass of the donor star (and
thus the chirp mass of the system to within a factor of $\sim$2) as
function of period are fixed \citep[e.g.][]{vil71,fau71}.
%However, both the mass transfer,
%as the donor mass at given period (and thus the chirp mass to within a
%factor of 2 or so) are determined only by the mass -- radius relation
%of the donor white dwarf. Because of the properties of the Roche-lobe
%radius and Keplers law the orbital period of a mass-transferring binary
%depends only on the mass -- radius relation of the donors star
%\citep[e.g.][]{ver97}. The mass transfer rate is a function of the
%loss of angular momentum due to GWR and the change of donor radius
%upon mass loss \citep[$\zeta_2$, e.g.][]{mns02}. 
Thus a measurement of $\dot{f}$ can test mass--radius models, which is
especially interesting because more realistic mass--radius relations
for low-mass white dwarfs become available \citep[e.g.][]{bil02}. In
Table~\ref{tab:chirp} we show the number of systems for which we can
measure the frequency change as function of the minimum phase shift.

Finally it would be very interesting to measure the second derivative
of the frequency, as that gives a direct check whether the change in
frequency is due to GWR only, or whether other effects (like tidal
torques etc.) are important \citep[][Phinney, private
communication]{wh98}. For pure GW losses we have\footnote{Note that
  \citet{wh98} give 5/3 as numerical factor, which is appropriate for
  the corresponding equation for the second derivative of the orbital
  period.}
\begin{equation}
\ddot{f} = \frac{11}{3} \, \frac{\dot{f}^2}{f}.
\end{equation}
As can be seen from Table~\ref{tab:chirp}, the number of systems for
which the second derivative can be measured is small, even for a more
optimistic assumption of a total \textit{LISA} operation time of 5 yr.
For none of the AM~CVn systems we expect its second derivatives to be
measurable. This is due to the fact that the AM~CVn systems form from
pairs of low-mass ($M \la 0.5 \msun$) white dwarfs
\citep[e.g.][]{npv+00} which have larger radii and thus do not
penetrate into the highest frequency range where $\ddot{f}$ is large
(see Fig.~\ref{fig:fh}). In this range there are only massive ($M > 1
\msun$) detached double white dwarf pairs that merge rather than
become AM~CVn systems.

\begin{table}
\begin{center}
\begin{tabular}{lrr|rr}
\hline
$\Delta \Phi_{\rm min}$ & \multicolumn{2}{c}{Detached systems} & \multicolumn{2}{c}{AM~CVn systems}
 \\
 & $\dot{f}$ & $\ddot{f}$  & $\dot{f}$ & $\ddot{f}$ \\ \hline
$\pi/2$ & 145 & 0 & 9 & 0 \\
$\pi/4$ & 313 & 0 & 22 & 0 \\
$\pi/8$ & 636 & 1 & 56 & 0 \\
$\pi/16$ &  1163 & 1 & 108 & 0 \\
$\pi/32$ & 1935 & 2 & 201 & 0 \\
$\pi/32$, $T$ = 5 yr & 10094 & 23 &  3416 & 0 \\ \hline
\end{tabular}
\end{center}
\caption[]{Number of systems for which $\dot{f}$ or $\ddot{f}$ can be
  measured as function of the minimum phase shift needed for
  detection. The duration of the \textit{LISA} mission is assumed to
  be 1 year, except for the last row, where a mission time of 5 yr is assumed.}
\label{tab:chirp}
\end{table}

\section{Results II: Short-period AM CV\lowercase{n} systems detectable by \textit{LISA}
  with optical and/or X-ray counterparts}\label{counterparts}

\begin{figure*}
  \resizebox{2\columnwidth}{!}{\includegraphics[angle=-90,clip]{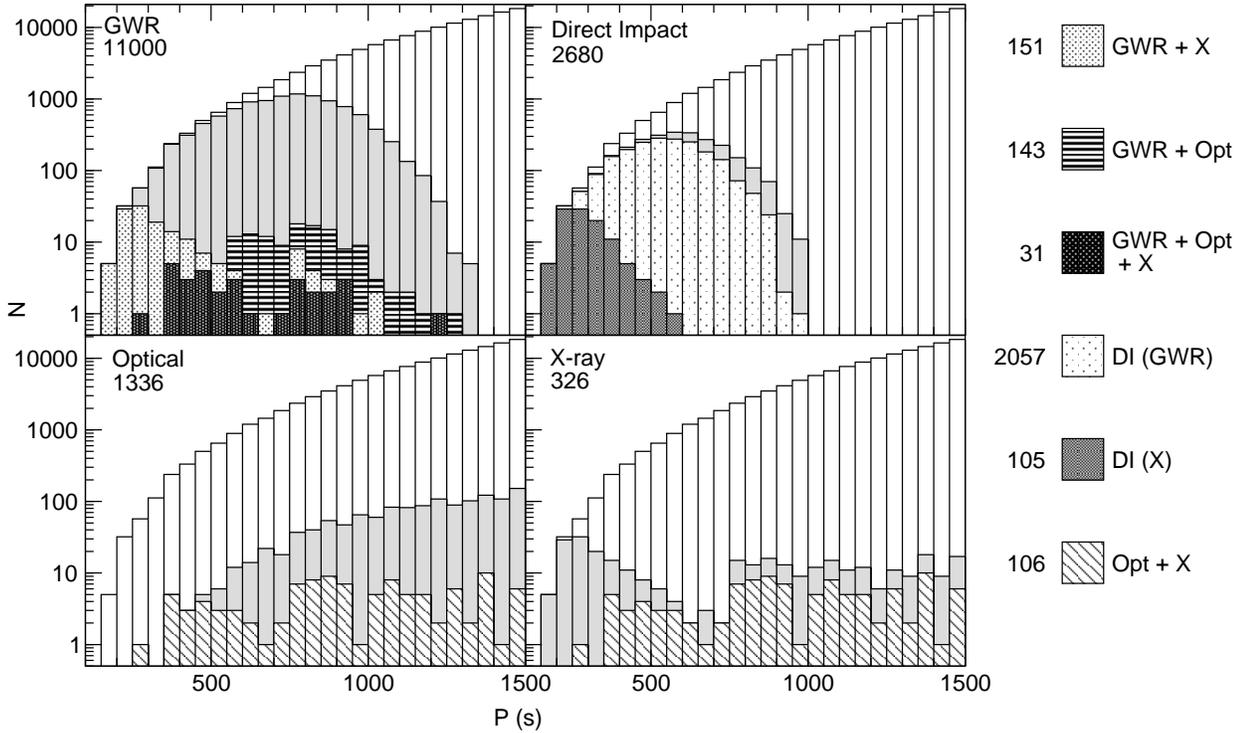}}

\caption[]{Histograms of the population of short-period AM~CVn
  systems, subdivided in different types. In each panel we show the
  \emph{total} AM~CVn population as the white histogram. In the top
  left panel we show the systems that can be resolved by \textit{LISA}
  in grey and subdivide these in the ones that have optical
  counterparts (GWR + Opt), X-ray counterparts (GWR + X) and both (GWR
  + Opt + X).  Similarly the top right panel shows the population that
  is in the direct impact phase of accretion in grey and we subdivide
  that population in GWR and X-ray sources. The bottom two panels show
  (again in grey) the populations that are detectable in the optical
  band (left panel) and the X-ray band (right panel). The distribution
  of sources detectable in both the optical and X-ray band is shown by
  the hatched histogram in both lower panels (Opt + X). See also
  Table~\ref{tab:resolved}.}
\label{fig:hist_all}
\end{figure*}

In Figure~\ref{fig:hist_all} we show the number distributions vs.
orbital periods 
%(up to 1500 s to show the transition to the longer period systems 
%discussed in \citetalias{npv+00}) 
for \textit{LISA} sources that have optical and/or X-ray counterparts,
direct impact systems that are GWR and X-ray sources, and the total
number of systems that are optical and/or X-ray sources (see also
Table~\ref{tab:resolved}). For comparison, we indicate also the
distribution of the total number of AM~CVn systems with $P < 1500$ s
as the white histogram in each panel.  The systems with different
properties are marked by different hatching styles. We will discuss
these groups now in more detail.

\textbf{Resolved \textit{LISA} sources (GWR)}\\
The top left panel shows the 11,000 systems that are expected to be
resolved by \textit{LISA} (grey histogram, also shown in
Fig.~\ref{fig:fh}), most of which do not have optical or X-ray
counterparts. There are 151 short-period systems that have X-ray
counterparts (GWR+X) and 143 systems, mainly with periods between 500
and 1000 s, that have optical counterparts (GWR + Opt).  Only 31
systems have both optical and X-ray counterparts (GWR + Opt + X).

\textbf{Direct impact accretors}\\
The top right panel shows the 2680 systems that are in the direct
impact phase of accretion (grey histogram, as in
Fig.~\ref{fig:histP}). Most of these systems (2057) are expected to be
resolved by \textit{LISA} [DI (GWR)], while at the shortest periods there are
105 X-ray systems that are in the direct impact phase [DI (X)].

\textbf{Optical sources}\\
The bottom left panel shows (in grey) that for periods below 25
minutes there are 1336 systems that have $V <$ 20 mag (grey
histogram). These are the systems shown in the bottom panel of
Fig.~\ref{fig:Pd}, and are the local population of AM~CVn systems.
Some (106) of these systems are also X-ray systems (Opt + X) and are
shown as the filled symbols in the top panel of Fig.~\ref{fig:Pd}.
\rev{Many of the short period systems will show eclipses \citep[even
  more than the detached systems which are discussed in][]{cfs03}
  which would compliment the GWR data and provide information on the
  radii of the components and the inclinations of the systems. E.g. a
  crude estimate for a typical system with initial parameters
  $0.25+0.6 \msun$ gives a probability for eclipsing the accreting
  star of about 30 per cent at $P=1000 s$, and even higher for
  eclipsing (part of) the accretion disc.}

\textbf{X-ray sources}\\
Finally, the bottom right panel shows the 326 X-ray sources that are
also shown in the top panel of Fig.~\ref{fig:Pd}. The majority is only
detectable in X-rays and many are in the direct impact phase.
\rev{Only a few direct impact systems are detectable in the optical
  band, by the radiation from the cooling white dwarf (the filled
  circles in Fig.~\ref{fig:Pd})}. At longer periods we expect
detectable X-ray emission from the \rev{boundary layers} of systems
formed from helium stars, some of which are also detectable in the
$V$-band (Opt + X).  \rev{These X-ray sources in the \textit{ROSAT}
  band in our model are concentrated in the Galactic plane, especially
  towards the Galactic centre. There are 193 systems (60 per cent of
  the total) in an area between Galactic longitudes $-$50 and +50
  degrees and Galactic latitude between $-$10 and +10 degrees. The
  \textit{ROSAT} all-sky
  survey\footnote{http://www.xray.mpe.mpg.de/rosat/survey/rass-fsc/}
  found $\sim$4500 sources in this region, so the short-period AM CVn
  systems can only account for a small fraction of these sources.}

\vspace*{0.3cm}
It is clear from Fig.~\ref{fig:hist_all} that \textit{LISA} is
expected to be very effective in detecting the short-period AM~CVn
systems. A large fraction of these will be in the direct impact phase.
At short periods, a significant fraction of the \textit{LISA} sources
might be detectable in X-rays, \rev{while at longer periods a small
  fraction might be optical sources.}
%, most of which are expected to be in
%direct impact phase.  Towards longer periods the systems detectable in
%the optical band take over. 
%%In the period range between
%%400 and 1000 s there is an interesting number of systems detectable by
%%\textit{LISA} and in the optical band.

\textit{LISA} sources with optical and/or X-ray counterparts are
important, as \textit{LISA} will measure a combination of all the
parameters that determine the GWR signal \citep[frequency, chirp mass,
distance, position in the sky and inclination angle, e.g.][]{hel03},
so if some of these parameters (period, position) can be obtained from
optical or X-ray observations the other parameters can be determined
with higher accuracy. This is particularly interesting for the
distances, inclinations and masses of the systems, which are very
difficult to measure with other methods.

\begin{table}
\begin{center}
\begin{tabular}{lrr}
\hline
 & total number & in direct impact phase \\
\hline
AM~CVn     & 138679      &  2680 \\
GWR        & 11000       &  2057  \\
Optical    & 1336        &  7     \\
X-ray      & 326         &  105   \\
GWR+Opt    & 143         &  3     \\
GWR+X      & 151         &  103   \\
GWR+Opt+X  & 31          &  3     \\
Opt+X      & 106         &  3     \\
% Gamma = 1.5:
%AM~CVn     & 200000      &  4000  \\
%GWR        & 11500       &  3000  \\
%Optical    & 2000        &  17    \\
%X-ray      & 500         &  100   \\
%GWR+Opt    & 160         &  12    \\
%GWR+X      & 160         &  100   \\
%GWR+Opt+X  & 30          &  6     \\
%Opt+X      & 190         &  6     \\
\hline
\end{tabular}
\end{center}
\caption[]{The number of short period ($P < 1500$ s) AM~CVn systems with different properties. The
  number of systems with each property that is in the direct impact
  phase is given in the second column. See also
  Fig.~\ref{fig:hist_all} for the distribution of these systems over
  orbital period.}
\label{tab:resolved}
\end{table}

\section{Discussion}\label{discussion}

We stress here that AM~CVn systems form from binaries with a rather
small range of initial parameters and thus that their birth rate is
very sensitive to small changes in assumptions in binary population
synthesis calculations. One has to keep in mind that a lot of these
assumptions are badly constrained, which means that the number of
progenitor systems is rather uncertain \citep[e.g.][]{nyp+00}. On top
of that, there are uncertainties connected with the formation of
AM~CVn systems from their immediate progenitors as discussed in
Sect.~\ref{popintro} and \citetalias{npv+00}, making the AM~CVn birth
rates uncertain by at least an order of magnitude.  The models \rev{
  we use} for the X-ray and optical emission are very simplified as
well. The interstellar absorption model might very well underestimate
the X-ray absorption for the systems in the Galactic centre, which
comprise a large fraction of the systems that are detectable only in
X-rays (pluses in Fig.~\ref{fig:Pd}). We furthermore recall that a
number of (possibly important) effects like irradiation and
compressional heating are not taken into account in our simple
emission models.

\rev{For comparison we also computed the results for X-ray detectors
  sensitive in the range 0.1 -- 15 keV (similar to the
  \textit{Chandra} and \textit{XMM} bands). Because most of our
  spectra are rather soft, the flux ($\mathcal{F}$) in this band is
  generally not much larger than in the \textit{ROSAT} band: 80 per
  cent has $\mathcal{F}_{\rm XMM}/\mathcal{F}_{\rm ROSAT} < $ 1.5; 96
  per cent has $\mathcal{F}_{\rm XMM}/\mathcal{F}_{\rm ROSAT} < 3$.
  However, the improved sensitivity of \textit{Chandra} and
  \textit{XMM} obviously will have an impact.  For instance the
  \citet{mbb+03} \textit{Chandra} observations of the Galactic Centre
  have a completeness limit of 3 10$^{-15}$ erg~cm$^{-2}$ s$^{-1}$,
  almost two orders of magnitude deeper than our assumed
  \textit{ROSAT} limit.  The number of X-ray sources in the 0.1 -- 15
  keV band detectable down to $10^{-14}$ erg~cm$^{-2}$~s$^{-1}$ is 644
  and 1085 down to $10^{-15}$ erg~cm$^{-2}$~s$^{-1}$. In the
  \textit{Chandra} mosaic image of the Galactic centre \citep{wgl02},
  roughly down to $10^{-14}$ erg~cm$^{-2}$~s$^{-1}$ there are
  $\sim$1000 point sources, while our model gives 16 X-ray systems in
  this region. In the tiny \citet{mbb+03} region, where they identify
  more than 2000 sources, we predict 3 AM~CVn systems with X-ray flux
  above $10^{-15}$ erg~cm$^{-2}$~s$^{-1}$.}

As briefly discussed in Sect.~\ref{AMCVnpop}, we did not include
AM~CVn systems formed like CVs in our models. The reason is that for
the production of the short period systems, which we consider in this
paper, this channel is not effective. An evolutionary track for a
donor with an initial mass of 0.96 \msun and a hydrogen-exhausted core
$\sim 0.05\,M_\odot$ \citep[sequence nr. 4 in][]{fe89} reaches a
minimum period of 8.4 minutes and spends 3.8 10$^{7}$ yr in the 25 --
8.4 minute period range (A.V.  Fedorova, private communication).
Taking this time as typical, and assuming a birth rate of 6.4
10$^{-5}$ yr$^{-1}$ for systems evolving to $P < 25$\,minutes, as in
the standard model of \citet{phr01}, we expect 2400 post-CV
ultra-compact systems in the Galaxy, compared to the 140,000 from the
channels discussed here.  However, if these channels are ineffective,
e.g., due to the instability of the mass transfer and/or destruction
of accretors by edge-lit detonations \citepalias{npv+00}, the AM~CVn
systems from CVs may become more important even at short periods.

%we found
%that number of systems formed through this channel with our
%assumptions about binary evolution is low and secondly they will
%hardly contribute to the short period systems, as the vast majority
%has a period minimum above 10 minutes. Defining possible AM~CVn
%systems as those with a minimum period below 1 hour, we find from the
%appendix of \citet{phr01} that one needs a white dwarf with a donor
%between 1 and 1.4 \msun. The birth rate for such systems in our
%calculations is only 3.5 $10^{-5}$ yr$^{-1}$. This is most likely due
%to the higher common envelope efficiency we use. In our model the
%birth rate of these systems was higher in the past (2 $10^{-4}$
%yr$^{-1}$), comparable to the current birth rate from helium stars
%(Table~\ref{tab:br_number}) but the birth rate from helium star
%binaries in our models at that time was almost an order of magnitude
%larger as well. So in our models, the number of systems from the CV
%channel is small. However, as mentioned in the previous section, one
%has to keep in mind the uncertainty in (especially the common
%envelope) parameters.

\section{Conclusion}\label{conclusion}

We discussed a model for the Galactic population of short-period
AM~CVn systems and the uncertainties in this model. We find a
  total population of AM~CVn systems in the Galaxy with periods
  shorter than 1500 s of $\sim$140,000 systems. We estimate that the
final uncertainty on this number is at least a factor of 10. The model
presented here gives an optimistic estimate of the total number of
systems, i.e. all quoted numbers could be (substantially)
smaller. We calculated the optical, X-ray and gravitational wave
signals that could be detectable, using simple emission models and
selection effects, and discussed the overlap between the
subpopulations which may be discovered by different detection methods.

In this \rev{optimistic} model we find that there are a few systems
expected that are in the direct impact phase and are detectable as
optical and X-ray sources, \rev{where the optical emission comes from
  the cooling of the (possibly heated) white dwarfs}.  Thus if RX
J0806 and V407 Vul turn out to be short period AM~CVn systems, that
would argue for an optimistic model \rev{as in a pessimistic model
  there would be no such systems that are observable}. \citet{str03}
and \citet{hrw+03} have reported a shortening of the periods in V407
Vul and RX J0826, arguing against this interpretation \rev{despite the
  fact that these stars might be helium-donor or CV systems before the
  period minimum (see also \citet{ww03} who argue that there is not
  sufficient evidence for a shortening of the period in RX J0806).}

At the observed periods of RX J0806, V407 Vul and ES Cet, the fraction
of systems that is in the direct impact phase is 80, 40 and 30 per
cent respectively. The total number of direct impact systems in the
Galaxy is 2680. We further find that there are 220 (very) short period
AM~CVn systems that are only detectable in the X-ray band, \rev{some}
with orbital periods as short as 3 minutes.

We conclude that the future gravitational wave detector \textit{LISA}
is expected to discover thousands of short-period AM~CVn systems,
about ten per cent of the total number of short-period systems in the
Galaxy. \rev{The number of systems with periods below 1500 s that is
  detectable in the optical band is 1336 (about one per cent of the
  total). Only 326 systems (0.25 per cent) are detectable in the ROSAT
  X-ray band.  For the typical sensitivities of \textit{Chandra} and
  XMM, the number of detectable sources over the whole sky is $\ga$
  1000, still only a tiny fraction of all X-ray sources.}

For many of the \textit{LISA} sources the change in frequency can be
measured, which would \rev{provide} a test of the mass -- radius
relation for the donor stars in these systems.  \textit{LISA} is
expected to detect a similar number of detached double white dwarfs,
for many of which the frequency change can be determined, yielding a
direct measurement of the distances to these sources.  Maybe for a few
the second derivative of the frequency can be measured, which would
give a direct test whether the frequency change is purely due to GWR,
or whether tidal effects play a role.

%Below 1000 s, we predict similar numbers of systems
%($\sim$200) detectable in the X-ray or optical band.  

Our results indicate that many short-period AM~CVn systems may be
detectable in more than one way. A total of 106 systems is detectable
in the optical and X-ray bands. For the systems that are detectable
with \textit{LISA}, about 300 (3 per cent) have optical and/or X-ray
counterparts, giving the opportunity to independently determine the
position and orbital period.\rev{ This leads to a better}
determination of the other parameters that can be derived from the GWR
signal (i.e.  inclination and masses) and would lead to unprecedented
\rev{accuracy in the study of} these extreme binaries.

\section*{Acknowledgments}

\rev{We thank the referee for comments that improved the paper} We
thank Samuel Boissier for providing the table of star formation rates
and helpful discussion about the Galactic model. GN is supported by
PPARC. LRY is supported by RFBR grant no.  03-02-16254 and Federal
Science and Technology Program ``Astronomy'' (contract no.
40.022.11.1102). SPZ is supported by the KNAW and NWO.

\bibliographystyle{NBmn}
\bibliography{journals,binaries}

\label{lastpage}

\appendix

\section{White dwarf cooling}\label{appendix}

\begin{figure}
  \resizebox{\columnwidth}{!}{\includegraphics[angle=-90,clip]{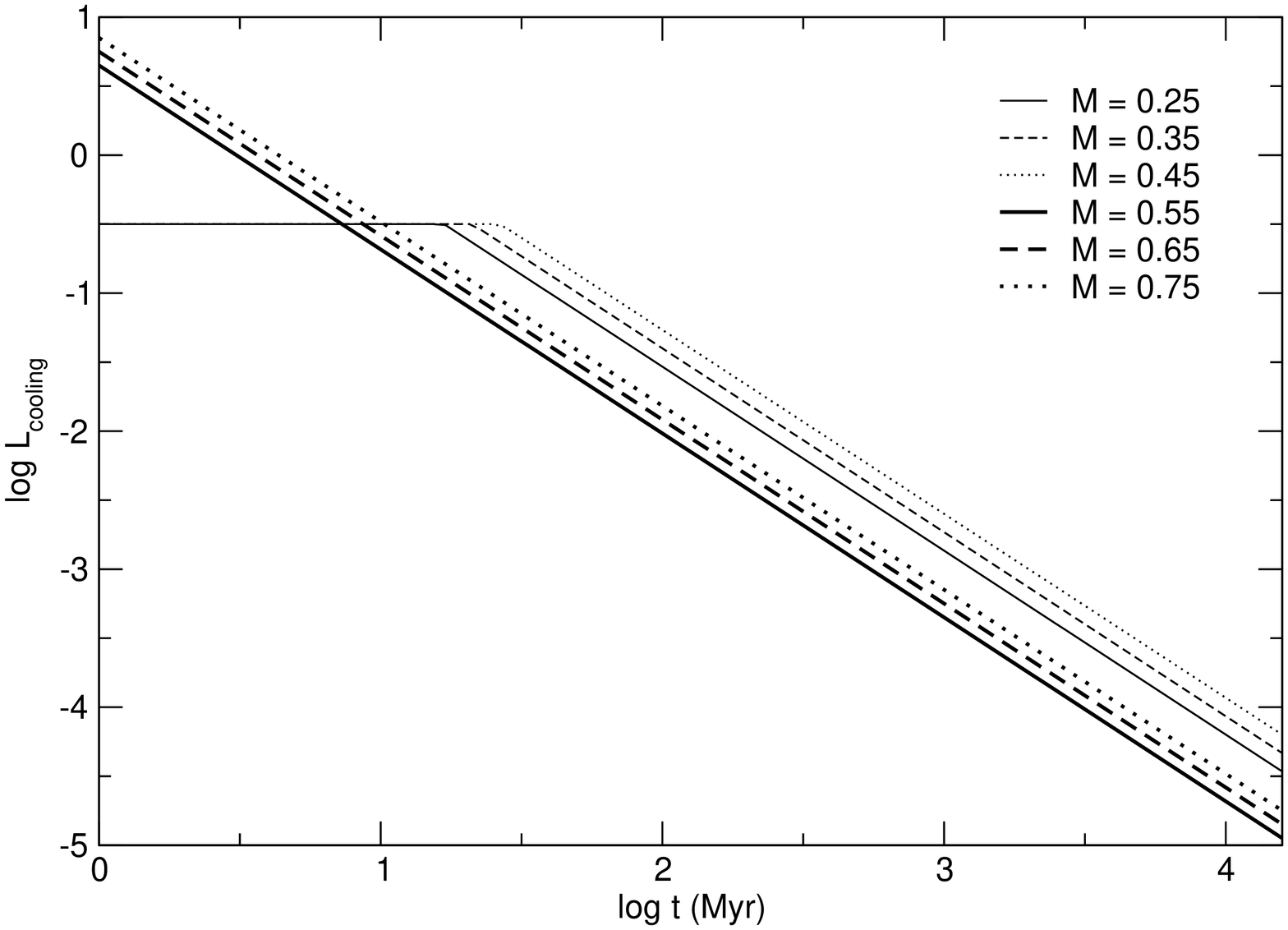}}
\caption[]{Cooling curves of white dwarfs used in this paper. The
  curves are a rough translation of Fig. 10 of \citet{han99}.}
\label{fig:wd_cooling}
\end{figure}

We model the cooling of white dwarf based on the curves of
\citet{han99}. We abandoned the previous use of the \citet{dsb+98}
models, since their models with thick hydrogen envelopes cool slower
than observations of detached double white dwarfs suggest and their
thick hydrogen envelopes actually may be lost in hydrogen shell
flashes \citep[e.g.][]{nyp+00}.

Our simple models have carbon-oxygen white dwarfs that cool faster than 
helium white dwarfs, but within the groups the most massive objects
are brighter at given age. We model the curves as (see
Fig.~\ref{fig:wd_cooling}):
\begin{equation}
\log L = L_{\rm max} - 1.33 \, \log (t/10^6 {\rm yr}),
\end{equation}
where $L_{\rm max}$ is a linear fit given by
\begin{equation}
L_{\rm max} = 1 - (0.9 - \; M_{\rm WD}) \quad \quad \quad  \quad \quad \mbox{ for } 
                     M_{\rm WD} > 0.5 \msun,
\end{equation}
and
\begin{equation}
L_{\rm max} = 1.4 - 1.33 (0.45 - \; M_{\rm WD}) \quad \mbox{ for } 
                     M_{\rm WD} < 0.5 \msun
\end{equation}
(mass and luminosity in solar units).  For white dwarf masses below
0.5 \msun the luminosity is constrained to be below
$\log\,L/L_{\sun}\,=\,-0.5$, for more massive white dwarfs below $\log
L/L_{\sun} = 2$.

\end{document}